\documentclass[12pt,a4paper]{article}
\usepackage{amssymb}
\usepackage{amsmath}
\RequirePackage{graphicx}
\RequirePackage{mathptmx}      
\RequirePackage{flushend}
\RequirePackage[numbers,sort&compress]{natbib}
\usepackage[colorlinks]{hyperref}
\hypersetup{
     breaklinks=true,
     pdfstartview={FitH},    % fits the width of the page to the window
    colorlinks=true,       % false: boxed links; true: colored links
    linkcolor=blue,          % color of internal links
    citecolor=red,        % color of links to bibliography
    filecolor=magenta,      % color of file links
    urlcolor=Green,           % color of external links
    anchorcolor=blue,      % Color for anchor text
    linktocpage=true
}
\usepackage{psfig}
\begin{document}
\begin{center}{\Large\bf A parametric reconstruction of the deceleration parameter}
\\[15mm]
Abdulla Al Mamon$^{a,}$\footnote{Present affiliation:\\ Manipal Centre for Natural Sciences, Manipal University,\\ Manipal-576104, Karnataka, India.}$^{,}$\footnote{E-mail : abdullaalmamon.rs@visva-bharati.ac.in, abdulla.mamon@manipal.edu}~and
Sudipta Das$^{a,}$\footnote{E-mail:  sudipta.das@visva-bharati.ac.in}\\
{\em $^{a}$Department of Physics, Visva-Bharati,\\
Santiniketan- 731235, ~India.}\\
[15mm]
\end{center}
\vspace{0.5cm}
\vspace{0.5cm}
\pagestyle{myheadings}
\newcommand{\be}{\begin{equation}}
\newcommand{\ee}{\end{equation}}
\newcommand{\bea}{\begin{eqnarray}}
\newcommand{\eea}{\end{eqnarray}}
\newcommand{\bc}{\begin{center}}
\newcommand{\ec}{\end{center}}
%%%%%%%%%%%%%%%%%%%%%%%%%%%%%%%%%%%%%%%%%%%%%%%%%%%%%%%
%%%%%%%%%%%%%%%%%%%%%%%%%%%%%%%%%%%%%%%%%%%%%%%%%%%%%%%
\begin{abstract}
The present work is based on a parametric reconstruction of the deceleration parameter $q(z)$ in a model for the spatially flat FRW universe filled with dark energy and non-relativistic matter. In cosmology, the parametric reconstruction technique deals with an attempt to build up a model by choosing some specific evolution scenario for a cosmological parameter and then estimate the values of the parameters with the help of different observational datasets. In this paper, we have proposed a logarithmic parametrization of $q(z)$ to probe the evolution history of the universe. Using the type Ia supernova (SNIa), baryon acoustic oscillation (BAO) and the cosmic microwave background (CMB) datasets, the constraints on the arbitrary model parameters $q_{0}$ and $q_{1}$ are obtained (within  $1\sigma$ and $2\sigma$ confidence limits) by $\chi^{2}$-minimization technique. We have then reconstructed the deceleration parameter, the total EoS parameter $\omega_{tot}$, the jerk parameter and have compared the reconstructed results of $q(z)$ with other well-known parametrizations of $q(z)$. We have also shown that two model selection criteria (namely, Akaike information criterion and Bayesian Information Criterion) provide the clear indication that our reconstructed model is well consistent with other popular models.
\end{abstract} 
%%%%%%%%%%%%%%%%%%%%%%%%%%%%%%%%%%%%%%%%%%%%%%%%%%%%%%%%%%%%%%%%%%%%%%%%%%%%%
\section{Introduction}
Recent observations strongly suggest that the universe is undergoing an accelerated expansion in the present epoch \cite{accsa1,accsa2}. The matter content responsible for such a certain stage of evolution of the universe is popularly referred to as ``dark energy" (DE), which accounts for almost $70\%$ of the current energy budget of the universe. In this regard, various DE models have been proposed to match with recent observed data and the $\Lambda$CDM model is the most simplest one in this series. But, this model suffers from some other problems which are known as the ``{\it fine tuning}" problem \cite{ccpsa1}, ``{\it coincidence}" problem \cite{ccpsa2} and the ``{\it age}" problem \cite{accsarev}. To overcome these issues, it is quite natural to think for some alternative possibilities to explain the origin and nature of DE. Models with scalar fields (both the canonical and non-canonical scalar field) play a major role in current description of the evolution of the universe. Motivated by the scalar field theories, over the last decade, numerous DE models were explored which include quintessence, K-essence, phantom, tachyon, chaplygin gas and so on (for review see \cite{accsabk} and references therein). However, we do not yet have a concrete and satisfactory DE model.\\
%%%%%%%%%%%%%%%%%%%%%%%%%%%%%%%%%%%%%%%%%%%%%%%%%%%%%%%%%%%%%%%%%%%%%%%%%%%%
\par As mentioned before, the cosmological observations also indicate that the observed cosmic acceleration is a recent phenomenon. So, in the absence of DE or when its effect is subdominant, the same model should have decelerated phase in the early epoch of matter era to allow the formation of structure (as gravity holds matter together) in the universe. For this reason, a cosmological model requires both the decelerated and an accelerated phase of expansion to describe the whole evolution history of the universe. In this context, the deceleration parameter plays an important role, which is defined as 
\be
q=-\frac{a{\ddot{a}}}{{\dot{a}}^2}
\ee
where $a(t)$ is the scale factor of the universe. For accelerating universe, ${\ddot{a}}>0$, i.e., $q<0$ and vice-versa. The most popular way of achieving such scenario is to consider a parametrization for the deceleration parameter as a function of the scale factor ($a$) or the redshift ($z$) or the cosmic time ($t$) (see Refs. \cite{dp1qlog,dp2qlog,dp3qloggw,dp3qlog,dp4qlog,dp5qlog,dp6qlog,dp7qlog,dp8qlog,dp9qlog,
dp10qlog,dp11qlog,dp12qlog}). It should be noted that for most of these parametrizations, the $q$-parametrization diverges at far future and others are valid at low redshift (i.e., $z<<1$) \cite{dp3qlog,dp4qlog,dp6qlog,dp8qlog,dp10qlog,dp11qlog}. Also such parametric methods may mislead about the true nature of dark energy due to the assumed parametric form. The non-parametric method is advantageous than the parametric method in the literature, since it avoids parametrizing cosmological quantities and finds the evolution of our universe directly from observational data \cite{npas,nphol1,nphol2,nprgc,nprn}. However, such approaches also have drawbacks \cite{npdb}. Till now, there is no well motivated theoretical model of the universe which can describe the entire evolution history of the universe. So, it is reasonable to adopt a parametric approach to measure the transition from a decelerating to an accelerating phase of the universe. In addition to this, parametric approach also helps to improve efficiency of the future cosmological surveys. Motivated by these facts, in the present work, we have chosen a special form of deceleration parameter in such a way that $q(z)$ will provide the desired property for sign flip from a decelerating to an accelerating phase. The properties of this parametrization is mentioned in section 2. The constraints on the model parameters of our toy model have also been obtained using the SNIa, BAO and CMB datasets. We have also reconstructed the redshift evolutions of $q(z)$ and the total EoS parameter and have compared their evolution behaviors with other well-known models, such as $q\propto z$ \cite{dp2qlog}, $q\propto \frac{z}{1+z}$ \cite{dp4qlog,dp6qlog,dp8qlog,dp10qlog,dp11qlog} and $\Lambda$CDM, to study the different properties of this model. We have shown that for this choice of $q(z)$, the present model describes the evolution of the universe from an early decelerated phase (where matter dominates over DE) to an accelerated phase (where DE dominates over matter) at the current epoch for each dataset.\\
%%%%%%%%%%%%%%%%%%%%%%%%%%%%%%%%%%%%%%%%%%%%%%%%%%
\par Another important kinematical quantity, related to the expansion of the universe, is {\it jerk parameter} (the dimensionless third derivative of the scale factor $a(t)$ with respect to cosmic time $t$) which is defined as \cite{jerk1,jerk2,jerk3,jerk4}
\be
j=\frac{\frac{d^3a}{dt^3}}{aH^3} 
\ee
and in terms of $q$,
\be\label{eqjerk}
j={\left[q(2q+1) + (1+z)\frac{dq}{dz}\right]} 
\ee
which will be useful when the parametric form of $q(z)$ is given. A powerful feature
of the jerk parameter is that for the standard $\Lambda$CDM model $j= 1$ (constant) always, which provides a simple test for departure from the $\Lambda$CDM scenario. It deserves mention here that Sahni et al. \cite{jerkvs}  and Alam et al. \cite{jerkua} drew attention to the importance of $j$ for discriminating various models of dark energy, because any deviation from $1$ in the value of $j$ would favor dynamical dark energy model, instead of $\Lambda$CDM model. In this paper, we have also investigated the evolution of $j$ for the present parametrized model. 
%%%%%%%%%%%%%%%%%%%%%%%%%%%%%%%%%%%%%%%%%%%%%%%%%%%%%%%%%%%%%%%%%%%%%%%%%%%%%%%%% 
\par The present paper is organized as follows. The basic equations for the phenomenological DE model has been presented in the next section. We have then tried to obtain some accelerating solutions for this toy model using a specific parametrization of $q(z)$. In section \ref{resu}, we have obtained the observational constraints on this model parameters using various datasets and
presented our results. Finally, in section \ref{conclusa}, there is a brief conclusions about the results obtained in this work.
%%%%%%%%%%%%%%%%%%%%%%%%%%%%%%%%%%%%%%%%%%%%%%%%%%%%%%%%%%%%%%%%%%%%%%%%%
\section{Basic equations and solutions}
%%%%%%%%%%%%%%%%%%%%%%%%%%%%%%%%%%%%%%%%%%%%%%%%%%%%%%%%%%%%%%%%%%%%%%%
Let us consider a spatially flat, homogeneous and isotropic FRW universe 
\be
ds^2 = dt^2 - a^2(t)[dr^2 + r^2 d{\Omega}^2] 
\ee
composed of two perfect fluids, namely ordinary matter with negligible pressure and canonical scalar field (as a candidate of dark energy). In this case, the Einstein's field equations become (choosing $8\pi G=c=1$)
\be\label{eqfeqlog1}
3H^{2}=\rho_{m} + \rho_{\phi}
\ee
\be\label{eqfeqlog2}
2{\dot{H}} + 3H^{2}=-p_{\phi}
\ee
where $H=\frac{\dot{a}}{a}$ is the Hubble parameter and $\rho_{m}$ is the matter energy density. Here, $\rho_{\phi}$ and $p_{\phi}$ are the contributions of the scalar field to the energy density and pressure respectively and are given by
\bea
\rho_{\phi}=\frac{1}{2}{\dot{\phi}}^{2} + V(\phi)\\
p_{\phi}=\frac{1}{2}{\dot{\phi}}^{2} - V(\phi)
\eea
where, an overhead dot denotes a derivative with respect to the cosmic
time $t$ and $V(\phi)$ is the potential associated with the scalar field $\phi$.\\
Also, the conservation equations for the scalar field and matter field read as
\be\label{eqfeqlog3}
{\dot{\rho}}_{\phi} + 3H(\rho_{\phi}+p_{\phi})=0
\ee
\be\label{eqfeqlog4}
{\dot{\rho}}_{m} + 3H\rho_{m}=0
\ee
Now one can easily solve the above equation to find the energy density for the normal matter as
\be
\rho_{m}=\rho_{m0}(1+z)^{3}
\ee
where $\rho_{m0}$ is an integration constant (which represents the matter energy density at the present epoch) and $z$ is the redshift parameter, defined as $z=\frac{1}{a}-1$.\\
The corresponding {\it equation of state} (EoS) parameter is now given by
\be
\omega_{\phi}=\frac{p_{\phi}}{\rho_{\phi}}=-\frac{(2{\dot{H}}+3H^2)}{(3H^2 -\rho_{m})}
\ee
which further leads to
\be\label{eqwphinw}
\omega_{\phi}=\frac{2q -1}{3 - 3\Omega_{m0}(1+z)^{3} {\left(\frac{H_{0}}{H}\right)}^{2} }
\ee
where $\Omega_{m0}=\frac{\rho_{m0}}{3H^2_{0}}$ is the matter density parameter at the present epoch.\\
From equations (\ref{eqfeqlog1}) and (\ref{eqfeqlog2}), one can obtain
\be
\frac{\ddot{a}}{a}=-\frac{1}{6}{\left(\rho_{m}+\rho_{\phi}+3p_{\phi}\right)},
\ee
and the present model will provide acceleration (${\ddot{a}}>0$) only if
\be\label{eqweffac}
\omega_{tot}=\frac{p_{\phi}}{\rho_{m}+\rho_{\phi}}=-\frac{(2{\dot{H}} + 3H^{2})}{3H^{2}}<-\frac{1}{3}
\ee
where $\omega_{tot}$ denotes the effective or total EoS parameter.\\
%%%%%%%%%%%%%%%%%%%%%%%%%%%%%%%%%%%%%%%%%%%%%%
Now, out of equations (\ref{eqfeqlog1}), (\ref{eqfeqlog2}), (\ref{eqfeqlog3}) and (\ref{eqfeqlog4}), only three are independent equations with four unknown parameters, namely, $H$, $\rho_{m}$, $\phi$ and $V(\phi)$. Thus, in order to solve the system of equations we need an additional input.
%%%%%%%%%%%%%%%%%%%%%%%%%%%%%%%%%%%%%%%%%%%%%%%%%%%%%%%%%%%%%%%%%%%%%%%%%%%%%%%%%%%%%%%
\par It is well-known that the parametrization of the deceleration parameter $q$ plays an important role in describing the nature of the expansion rate of the universe. In general, $q$ can be parametrized as
\be\label{eqparaqm}
q(z)=q_{0} + q_{1}X(z) 
\ee
where $q_{0}$, $q_{1}$ are real numbers and $X(z)$ is a function of redshift $z$. In fact, various functional forms of $X(z)$ have been proposed in the literature \cite{dp1qlog,dp2qlog,dp3qloggw,dp3qlog,dp4qlog,dp5qlog,dp6qlog,dp7qlog,dp8qlog,dp9qlog,
dp10qlog,dp11qlog,dp12qlog}, which can provide a satisfactory solution to some of the cosmological problems. However, as mentioned earlier, some of these parametrizations lose their prediction capability regarding the future evolution of the universe) and others are valid for $z<<1$ only. In Ref. \cite{dp12qlog} the authors have considered a divergence-free parametrization of $q$ to study the whole expansion history of the universe. They have shown that such model is more consistent with the current observational constraints for some restrictions on model parameters. Hence, search is still on for an appropriate functional form of $q(z)$ that will fit well in dealing with cosmological challenges. Motivated by these facts, in this present work, we have proposed a parametrization of the deceleration parameter, which is given by
\be\label{eqdplog}
q(z)=q_{0} + q_{1}\left(\frac{\ln(N+z)}{1+z}-\kappa\right)
\ee
where $q_{0}$, $q_{1}$, $N$ and $\kappa$ are arbitrary model parameters. The system of equations is closed now. It is straight forward to see that $q(z)$ has the following limiting cases:
%%%%%%%%%%%%%%%%%%%%%%%%%%%%%%%%%%%%%%%%%%%%%%%
\bea\label{eqdploglc}
{q(z)} = \left\{\begin{array}{ll} q_{0}-q_{1}\kappa,&$for$\ z\rightarrow +\infty\hspace{1mm} ({\rm early~epoch}),\\\\
q_0 +q_{1}(\ln N -\kappa),\ \ \ \ \ \ \ \ \ \ &$for$\
z=0\hspace{1mm} ({\rm present~epoch})\\
\end{array}\right.
\eea
%%%%%%%%%%%%%%%%%%%%%%%%%%%%%%%%%%%%%%%%%%%%%%%%%%%
Clearly, one can realize the history of the cosmic evolution with this new parametrization. Similarly, at low redshift ($z<<1$), the form of $q(z)$ comes out to be
\be\label{eqlowz}
q(z) = q_{0} + q_{1}{\left(\frac{\ln N}{(1+z)}+\frac{z}{N(1+z)}-\kappa\right)}
\ee
Now, we try to reproduce different well-known forms of $q(z)$ from equation (\ref{eqlowz}) for various choices of $\kappa$.\\
i) For $\kappa=\ln N$, one can easily obtain the expression for $q(z)$ as
\be\label{eqcpl}
q(z) = q_{0} + {\tilde{q}}_{1}{\left(\frac{z}{1+z}\right)},\hspace{3mm} {\rm where}\hspace{1.5mm} {\tilde{q}}_{1} = q_{1}{\left(\frac{1}{N} - \ln N \right)}
\ee
which is similar to the well-known parametrization of $q(z)$ used by several authors \cite{dp4qlog,dp6qlog,dp8qlog,dp10qlog,dp11qlog}.\\
ii) For $\kappa=0$, $q(z)$ reduces to
\be\label{eqlz2}
q(z) = q_{0} + \frac{q_{1}z + q_{2}}{N(1+z)} 
\ee
where, $q_{2}=q_{1}N\ln N$. Note that the above form of the deceleration parameter is similar to the parametrization of $q(z) = \frac{1}{2} + \frac{q_{1}z + q_{2}}{(1+z)^{\epsilon}}$, where $\epsilon$ is an arbitrary constant. A similar form of deceleration parameter has been used by many authors with $\epsilon=2$ \cite{dp3qloggw,dp3qlog}, where they have shown that $q(z) = \frac{1}{2} + \frac{q_{1}z + q_{2}}{(1+z)^{2}}$.\\
iii) Similarly, for $\kappa=\frac{1}{N}$, we have obtained\\
\be\label{eqlz3}
q(z) = q_{0} + \frac{{q}^{*}_{1}}{1+z},\hspace{3mm} {\rm where}\hspace{1.5mm} {q}^{*}_{1} = \frac{q_{1}}{N}{\left(N\ln N -1\right)}
\ee
which is similar to the form of $q(z) = \frac{1}{2} + \frac{{q}_{1}}{(1+z)^{\epsilon}}$ for $\epsilon=1$ and appropriate choices of $q_{0}$, $q_{1}$ and $N$ \cite{dp10qlog}.\\ \\
%%%%%%%%%%%%%%%%%%%%%%%%%%%%%%%%%%%%%%%%%%%%%%%%%%%%%%%%%%%%
Therefore, the new parametrization of $q(z)$, given by equation (\ref{eqlowz}), covers a wide range of other popular theoretical models (as given in equations (\ref{eqcpl}), (\ref{eqlz2}) and (\ref{eqlz3})) for $z<<1$ and different choices of $\kappa$.\\
%%%%%%%%%%%%%%%%%%%%%%%%%%%%%%%%%%%%%%%%%%%%%%%%%%%
\par However, for our analysis, in the present work, we have considered the parametrization of $q(z)$ as  
\be\label{eqdplog2g}
q(z)=q_{0} + q_{1}\left(\frac{\ln(N+z)}{1+z}-\ln N\right),~~~~~N>1~~~~~~~~({\rm Model}~1)
\ee
which can be derived from equation (\ref{eqdplog}) by replacing $\kappa={\rm ln}N$. 
The reason for making this choice is that at $z=0$ (i.e., at present epoch), the second term in equation (\ref{eqdplog2g}) vanishes and $q_{0}$ provides the present value of $q(z)$.\\
%%%%%%%%%%%%%%%%%%%%%%%%%%%%%%%%%%%%%%%%%%%%%%%%%
\par In general, $q_{1}$ is another model parameter which characterizes the evolution of $q(z)$. To ensure the matter dominated epoch i.e., $q = \frac{1}{2}$ at the high $z$, one can constrain $q_{0}-q_{1}\ln N=\frac{1}{2}$ for this parametrization. In this case, the equation (\ref{eqdplog2g}) reduces to
\be
q(z)=q_{0} + \frac{2q_{0}-1}{2\ln N}\left(\frac{\ln(N+z)}{1+z}-\ln N\right)
\ee
In that case, the three parameter parametrization reduces to two parameter parametrization, namely $q_{0}$ and $N$. Since, we are interested to obtain a viable dynamical dark energy model and we can not ignore the importance of the two parameters $q_{0}$ and $q_{1}$ to probe the evolutionary history of the universe. For this reason, in this work, we have considered the two parameter ($q_{0},q_{1}$)-phase space with their proper physical interpretation. We can now easily constrain them using the available observational data, so that the above mentioned toy model can explain the present evolution of the universe more precisely.  \\
%%%%%%%%%%%%%%%%%%%%%%%%%%%%%%%%%%%%%%%%%%%%%%%%%%%%%%%%%%%%%%%%%%%%%%%%%%%%
The Hubble parameter and the deceleration parameter are related by the following equation
\be\label{eqhqlog}
H(z)=H_{0}{\rm exp}{\left(\int^{z}_{0} \frac{1+q(z^{\prime})}{1+z^{\prime}}dz^{\prime}\right)}
\ee
where $H_{0}$ is the present value of the Hubble parameter. With the help of equations (\ref{eqdplog2g}) and (\ref{eqhqlog}), we have obtained the expression for Hubble parameter $H$ as
\be\label{eqhlog}
H(z)=H_{0}N^{\frac{q_{1}N}{N-1}}(1+z)^{\alpha} (N+z)^{-\frac{q_{1}(N+z)}{(1+z)(N-1)}}
\ee
where $\alpha={\left(1+q_{0}+\frac{q_{1}}{N-1} -q_{1}\ln N\right)}$.\\ 
%%%%%%%%%%%%%%%%%%%%%%%%%%%%%%%%%%%%%%%%%%%%%
For this model, the expressions for the EoS parameter $\omega_{\phi}$ can be easily obtained (using equations (\ref{eqwphinw}), (\ref{eqdplog2g}) and (\ref{eqhlog})) as 
\be\label{eqwphilog}
\omega_{\phi}(z)=\frac{2q_{0} - 1 + 2q_{1}\left(\frac{\ln(N+z)}{1+z}-\ln N\right)}{3-3\Omega_{m0}N^{\frac{-2q_{1}N}{N-1}}(1+z)^{3-2\alpha} (N+z)^{\frac{2q_{1}(N+z)}{(1+z)(N-1)}}}
\ee
%%%%%%%%%%%%%%%%%%%%%%%%%%%%%%%%%%%%%%%%%%%%%%%%%%%%%%%%%%%%%%%%%%%%%%%%%%%%%%%%%%%%%%%%
It is evident from equation (\ref{eqwphilog}) that the EoS parameter $\omega_{\phi}(z)$ at low redshift reduces to 
\be
\omega_{\phi}(z)=\frac{2q_{0}-1}{(3-3\Omega_{m0})} + \frac{2{\tilde{q}}_{1}}{(3-3\Omega_{m0})}{\left(\frac{z}{1+z}\right)}
\ee
which is similar to the CPL parametrization of $\omega_{\phi}(z)$ given by $\omega_{\phi}(z)=\omega_{0} + \omega_{1}{\left(\frac{z}{1+z}\right)}$, and has been frequently used for many cosmological analysis \cite{cpl1,cpl2}.\\
%%%%%%%%%%%%%%%%%%%%%%%%%%%%%%%%%%%%%%%%%%%%%%%%%%%%%%%%%%%%%%%%%%%%%%%%%
\par Now, using equations (\ref{eqweffac}) and (\ref{eqhlog}), the total EoS parameter is obtained as
\bea\label{eqomtotal}
\omega_{tot}(z)=-1+\frac{2(1+z)}{3H}\frac{dH}{dz}= -1 + \frac{2}{3}\Big[1+q(z)\Big]\nonumber \\
=-\frac{1}{3}{\left[1-2q_{0}+2q_{1}{\rm ln}N - 2q_{1}\frac{{\rm ln}(N+z)}{1+z}\right]}
\eea
%%%%%%%%%%%%%%%%%%%%%%%%%%%%%%%%%%%%%%%%%%%%%%%%%%%%%%%%%%%%%%%%%%%%%%%%%%%%%%%%%
For the present model, the cosmic jerk parameter $j(z)$ can be obtained (using equations (\ref{eqjerk}) and (\ref{eqdplog2g})) as
\be\label{eqjerkz30}
j(z)=q_{0}+q_{1}{\left(\frac{1}{N+z}-\ln N\right)} + 2{\left[q_{0}+q_{1}{\left(\frac{\ln (N+z)}{1+z}-\ln N\right)}\right]}^2
\ee
with $j_{0}=j(z=0)=q_{0}+2q^2_{0} + \frac{q_{1}}{N}(1-N\ln N)$.\\
%%%%%%%%%%%%%%%%%%%%%%%%%%%%%%%%%%%%%%%%%%%%%%%%%%%%%%%%%%%%%%%%%%%%%%%%%%%%%%
\par In the remaining part of this paper, the possibility of having a transition of the expansion of the universe from a decelerated to an accelerated one is investigated. It is seen from equation (\ref{eqdplog2g}) that the functional form of $q(z)$ depends crucially on the values of the model parameters, namely, $q_{0}$, $q_{1}$ and $N$. So, in principle one can choose these parameters arbitrarily and study the functional behavior of $q(z)$ to confront with observational dataset. But, in this work, we first constrain the model parameters ($q_{0},q_{1}$) for some specific values of $N$ and using various observational datasets and with the best fit values obtained, we then try to reconstruct $q(z)$, $\omega_{tot}(z)$ and $j(z)$. We have also checked that the present analysis is consistent with the observational datasets for higher values of $N$ ($N>2$) as well. Interestingly, we have also found that for $N=2$, the logarithmic form of $q(z)$ (as given in equation (\ref{eqdplog2g})) is similar to the divergence-free parametrization of the dark energy EoS parameter \cite{moti}.\\
%%%%%%%%%%%%%%%%%%%%%%%%%%%%%%%%%%%%%%%%%%%%%%%%%%%%%%%%%%%%%%%%%%%%%%%%% 
\par For a comprehensive analysis, we have also compared our theoretical model with the following well-known models (for details, see sub-section \ref{resultsda}):\\ \\
i) Firstly, we have considered the linear redshift parametrization of
$q(z)$, which has the following functional form \cite{dp2qlog}
\be\label{eqmodel2}
q(z)=q_{0}+q_{1}z~~~~~~~~~~({\rm Model}~2)
\ee
where, $q_{0}$ and $q_{1}$ represent the present value and the first derivative of $q(z)$ respectively. In this case, $H(z)$ evolve as
\be
H(z)=H_{0}(1+z)^{(1+q_{0}-q_{1})}e^{q_{1}z}
\ee
ii) Secondly, we have considered a $q$-parametrization of the following
functional form \cite{dp4qlog,dp6qlog,dp8qlog,dp10qlog,dp11qlog} 
\be\label{eqmodel3}
q(z)=q_{0}+\frac{q_{1}z}{1+z}~~~~~~~~~~({\rm Model}~3)
\ee 
For this model, the Hubble parameter can be obtained as
\be
H(z)=H_{0}(1+z)^{(1+q_{0}+q_{1})}e^{-\frac{q_{1}z}{1+z}}
\ee
iii) Next, we have considered a flat $\Lambda$CDM model. The corresponding form of $q(z)$ is given by 
\be \label{eqmodellcdm}
q(z)=-1+\frac{3}{2{\left[1+\frac{\Omega_{\Lambda 0}}{\Omega_{m0}}(1+z)^{-3}\right]}}~~~~~~~(\Lambda {\rm CDM~model})
\ee
where, $\Omega_{\Lambda 0}+\Omega_{m0}=1$.
In this case, the solution for $H(z)$ can be obtained as
\be
H(z)=H_{0}[\Omega_{m0}(1+z)^{3} + (1-\Omega_{m0})]^{\frac{1}{2}}
\ee
%%%%%%%%%%%%%%%%%%%%%%%%%%%%%%%%%%%%%%%%%%%%%%%%%%%%%%%%%%%%%%%%%%%%%%%%
For each of these models (Models 1, 2 and 3), we have performed statistical analysis to constrain the parameters ($q_{0},q_{1}$) of the models and with the best fit values of these parameters, the evolution of various relevant cosmological parameters have been studied.
%%%%%%%%%%%%%%%%%%%%%%%%%%%%%%%%%%%%%%%%%%%%%%%%%%%%%%%%%%%%%%%
\section{Observational constraints and results}\label{resu}
%%%%%%%%%%%%%%%%%%%%%%%%%%%%%%%%%%%%%%%%%%%%%%%%%%%%%%%%%%%%%%%%%%%%%%%%%%
In this section, we have described the observational datasets and the statistical analysis method that will be used to put constraints on the various parameters of the model presented in the previous section and then discussed the results obtained in this analysis. In this work, we have used the recent observational datasets from the type Ia supernova (SNIa), baryon acoustic oscillation (BAO) and the cosmic microwave background (CMB) radiation observations.
%%%%%%%%%%%%%%%%%%%%%%%%%%%%%%%%%%%%%%%%%%%%%%%
\subsection{Type Ia supernova data}
%%%%%%%%%%%%%%%%%%%%%%%%%%%%%%%%%%%%%%%%%%%%%%%%
In the present work, the 31 binned distance modulus data sample of the recent joint lightcurve analysis has been utilized \cite{jla}. For this dataset, the $\chi^{2}$ is defined as (for more details see \cite{jlam})
\be
\chi^{2}_{SNIa}=A(p)-\frac{{B}^2(p)}{C}-\frac{2{\rm ln}10}{5C}B(p)-Q
\ee
where
\bea
A(p)=\sum_{\alpha,\beta}(\mu^{th}-\mu^{obs})_{\alpha}({\cal C}ov)^{-1}_{\alpha\beta}(\mu^{th}-\mu^{obs})_{\beta},\\
B(p)=\sum_{\alpha}(\mu^{th}-\mu^{obs})_{\alpha}\sum_{\beta}({\cal C}ov)^{-1}_{\alpha\beta},~~~~~~~~~~~~\\
C=\sum_{\alpha,\beta}({\cal C}ov)^{-1}_{\alpha\beta}~~~~~~~~~~~~~~~~~~~~~~~~~~~~~~~~~~~~~~~~
\eea
and the $``{\cal C}ov"$ is the $31\times31$ covarience matrix of the binned data sample. Here, $Q$ is a constant that does not depend on the model parameter $p$ and hence has been ignored. Also, $\mu^{th}$ and $\mu^{obs}$ represent the theoretical and observed distance modulus respectively.
%%%%%%%%%%%%%%%%%%%%%%%%%%%%%%%%%%%%%%%%%
\subsection{BAO/CMB data}
%%%%%%%%%%%%%%%%%%%%%%%%%%%%%%%%%%%%%%%%%%%%%%%
Next, we have considered baryon acoustic oscillation (BAO) \cite{bao1,bao2,bao2a,bao3} and cosmic microwave background (CMB) \cite{cmb} radiation measurement dataset to obtain the BAO/CMB constraints on the model parameters. For BAO data, the results from the WiggleZ Survey \cite{bao3}, SDSS Galaxy sample \cite{bao2,bao2a} and 6dF Galaxy Survey \cite{bao1} datasets have been adopted. Also, the CMB measurement considered is derived from the Planck2015 observations \cite{cmb}. One can look into Refs. \cite{mvdos}, where the required datasets are given in a tabular form. The details of methodology for obtaining the BAO/CMB constraints on model parameters is available in Ref. \cite{mvdos}, where the $\chi^2$ function is defined as
\be\label{eqchibaocmb}
\chi^2_{BAO/CMB} = X^{T}C^{-1}X
\ee
Here, $X$ and $C^{-1}$ are the transformation matrix and the inverse covariance matrix respectively \cite{mvdos}.
%%%%%%%%%%%%%%%%%%%%%%%%%%%%%%%%%%%%%%%%%%%%%%%%%%%%%%%%%%%%%%%%%%%%%%%%%%%%%%%%%%%
\par To constrain cosmological parameters from a joint analysis of the SNIa and BAO/CMB
datasets, we compute 
\be
\chi^{2}=\chi^2_{SNIa}+\chi^2_{BAO/CMB} 
\ee
The best-fit corresponds to the model parameters for which the above $\chi^{2}$ is minimum. In this work, we have minimized $\chi^{2}$ with respect to the model parameters $q_{i}$ (where, $i=0,1$) to estimate their best fit values. For this purpose, we have fixed the parameter $N$ to some constant value. From $\chi^{2}$, one can also compute the total likelihood function ${L}$, which is given by
\be
{L}(q_{i})=e^{-\frac{\chi^{2}_{T}(q_{i})}{2}} = {L}_{SNIa}(q_{i})\times {L}_{BAO/CMB}(q_{i})
\ee
where, ${L}_{SNIa}(q_{i})=e^{-\frac{\chi^{2}_{SNIa}(q_{i})}{2}}$ and ${L}_{BAO/CMB}(q_{i})=e^{-\frac{\chi^{2}_{BAO/CMB}(q_{i})}{2}}$ are the likelihood functions of the SNIa and BAO/CMB datasets respectively.
%%%%%%%%%%%%%%%%%%%%%%%%%%%%%%%%%%%%%%%%%%%%%%%%%%%%%%%%%%%%%%%
\subsection{Results of the data analysis}\label{resultsda}
%%%%%%%%%%%%%%%%%%%%%%%%%%%%%%%%%%%%%%%%%%%%%%%%%%%%%%%%%%%%%%%%%%%%%%%%%%%%%%%%%%
\par
\begin{figure}
\centering
\includegraphics[width=0.30\textwidth,height=0.20\textheight]{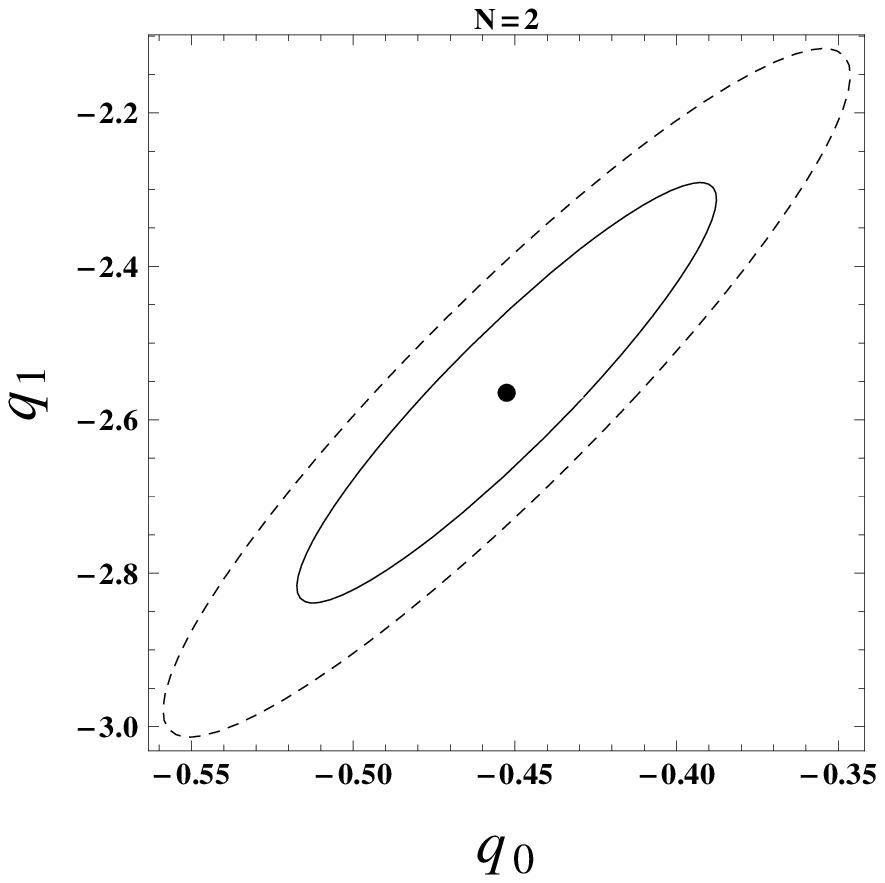}\hspace{2mm}
\includegraphics[width=0.30\textwidth,height=0.20\textheight]{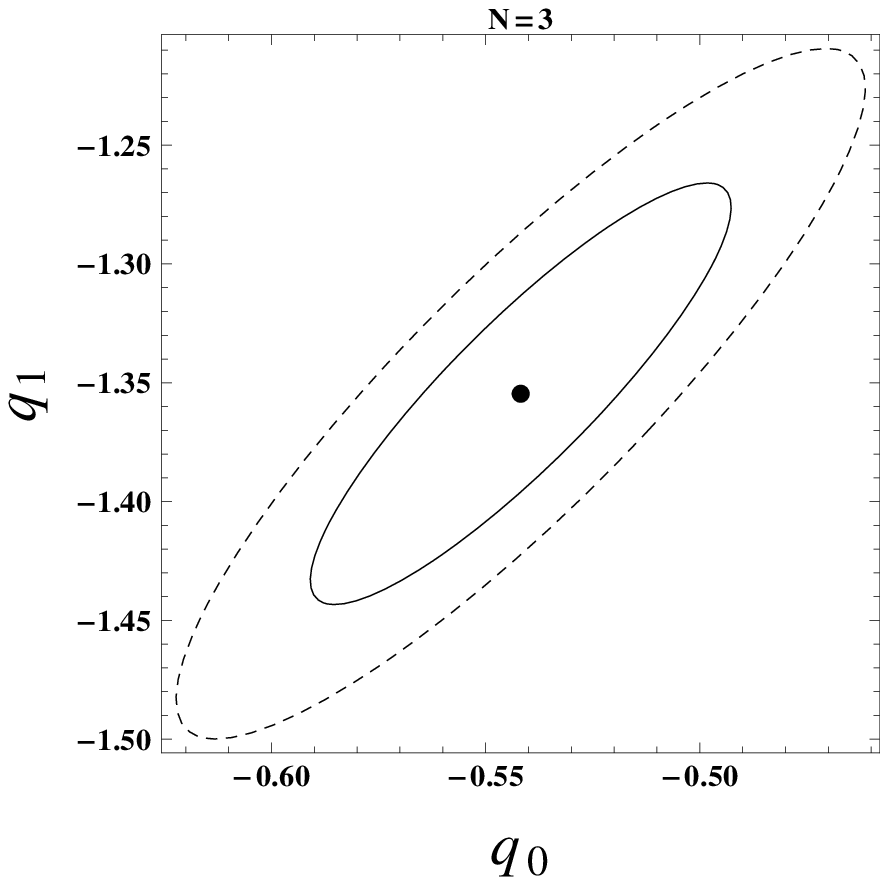}\\
\vspace{2mm}
\includegraphics[width=0.30\textwidth,height=0.20\textheight]{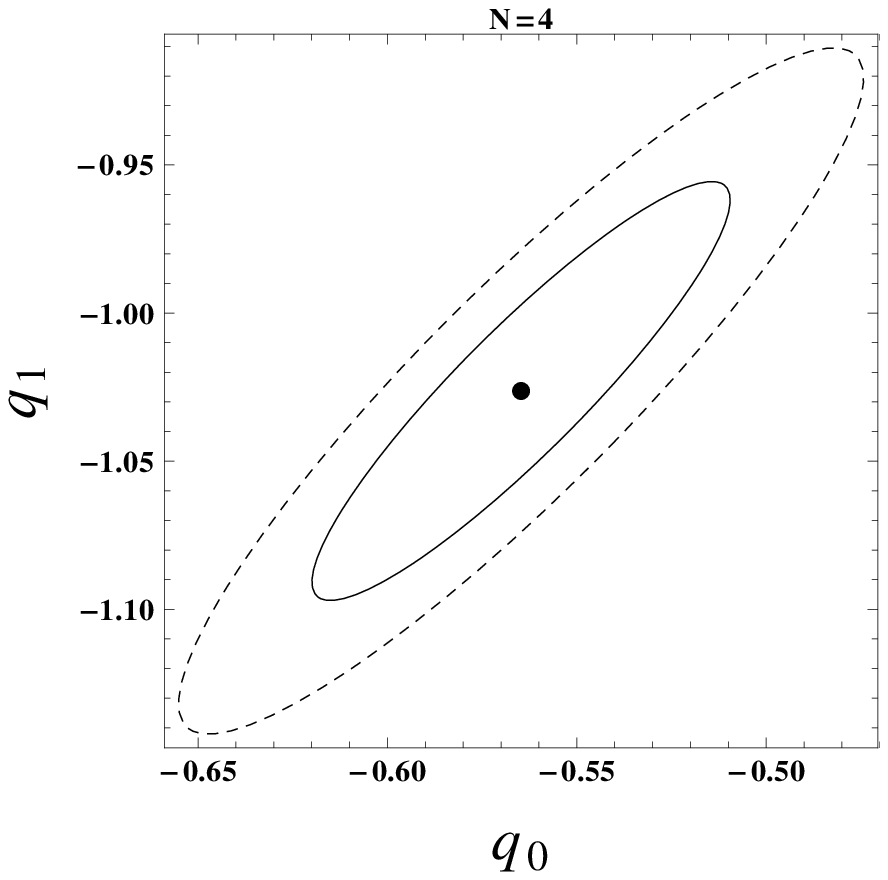}\hspace{2mm}
\includegraphics[width=0.30\textwidth,height=0.20\textheight]{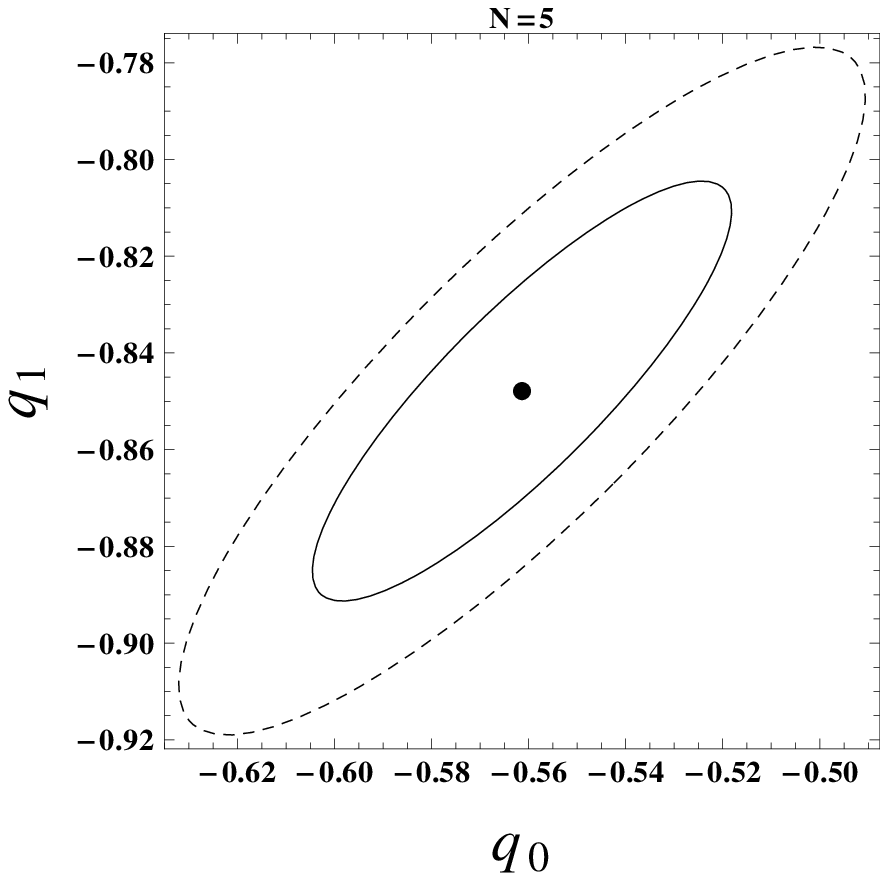}
\caption{\normalsize{\em $1\sigma$ (solid line) and $2\sigma$ (dashed line) confidence level contours in ($q_{0}$, $q_{1}$) plane have been shown for different choices of $N$. In each panel, the large dot indicates the best fit values of $q_{0}$ and $q_{1}$ arising from the analysis of SNIa+BAO/CMB dataset.}}
\label{figcontourqcsn}
\end{figure}

\begin{figure}
\centering
\includegraphics[width=0.30\textwidth,height=0.18\textheight]{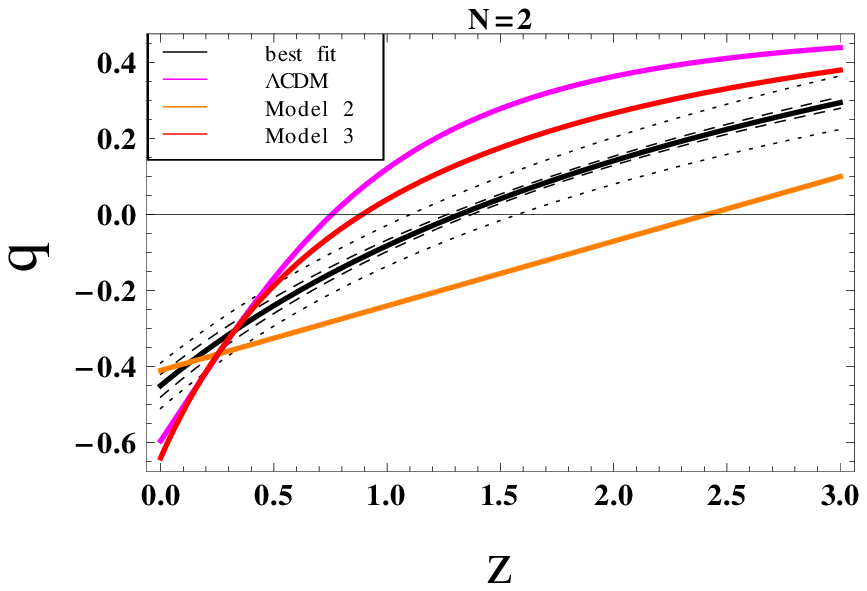}\hspace{2mm}
\includegraphics[width=0.30\textwidth,height=0.18\textheight]{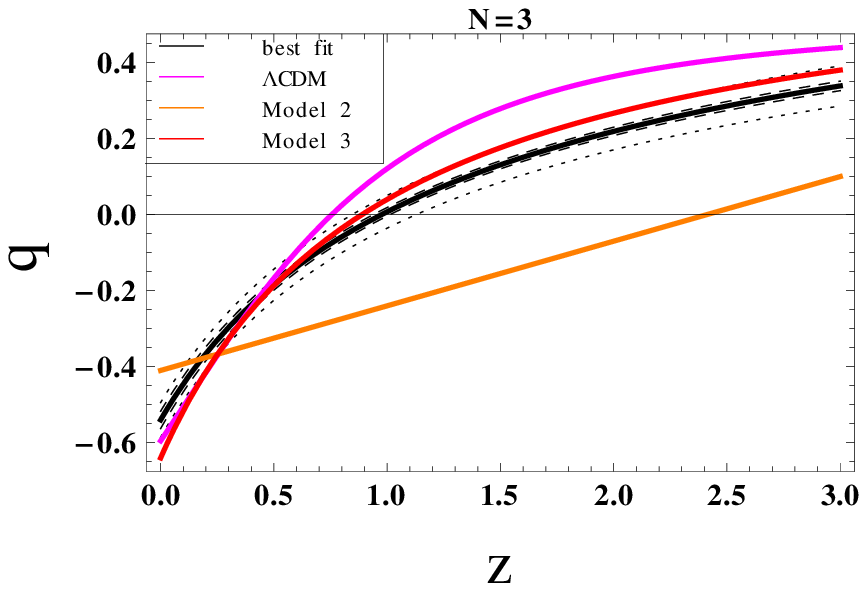}\\
\vspace{2mm}
\includegraphics[width=0.30\textwidth,height=0.18\textheight]{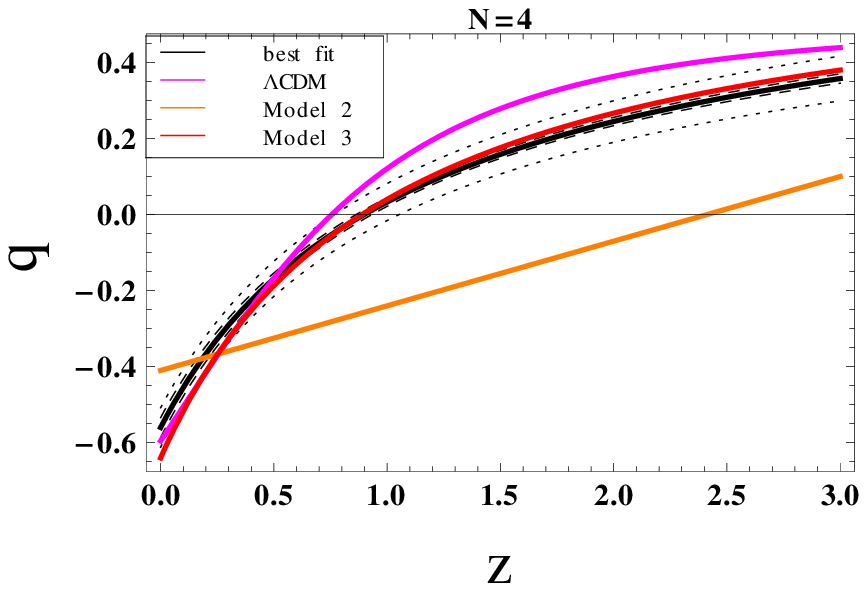}\hspace{2mm}
\includegraphics[width=0.30\textwidth,height=0.18\textheight]{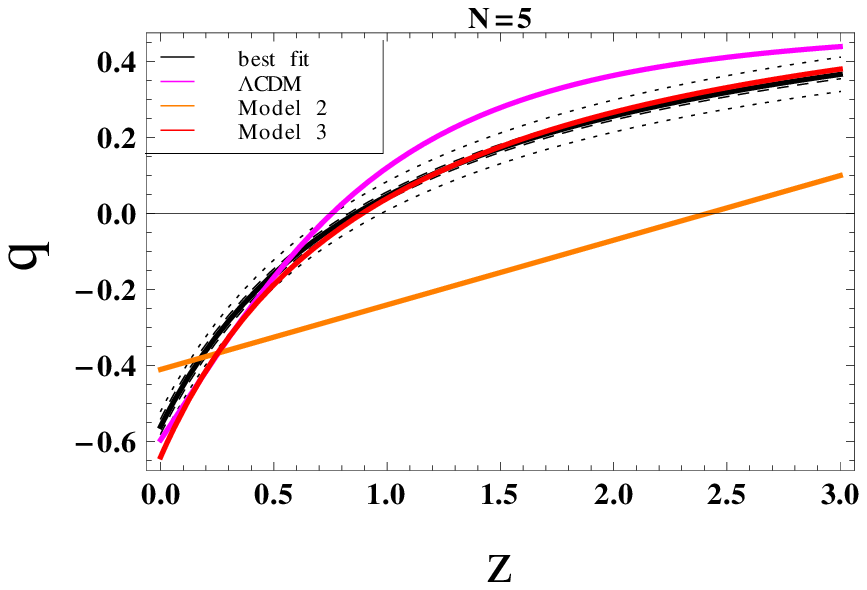}
\caption{\normalsize{\em The deceleration parameter $q(z)$ is reconstructed for the parametrization given by equation (\ref{eqdplog2g}) using different values of $N$, as indicated in the each panel. The central thick line (black) represents the best-fit curve, while the dashed and dotted contours represent the $1\sigma$ and $2\sigma$ confidence level respectively. In each panel, the orange, red and magenta lines represent the trajectory of $q$ for the Model 2, 3 and the $\Lambda$CDM model (with $\Omega_{\Lambda 0}=0.73$) respectively. Also, the horizontal thin line indicates $q(z) = 0$.}}
\label{figcontourqz}
\end{figure}

\begin{figure}
\centering
\includegraphics[width=0.32\textwidth,height=0.25\textheight]{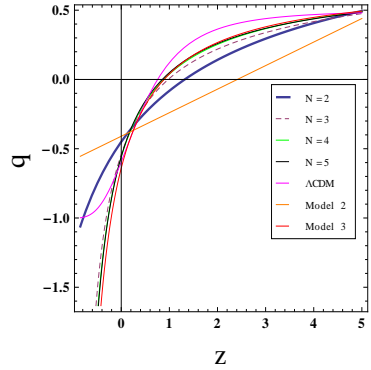}
\caption{\normalsize{\em Evolution of $q(z)$ is shown upto $z>-1$ for various models using the best-fit values of $q_{0}$ and $q_{1}$ arising from the joint analysis of SNIa+BAO/CMB dataset (see table \ref{table1} $\&$ \ref{table2}).}}
\label{figqfall}
\end{figure}

\begin{figure}
\centering
\includegraphics[width=0.32\textwidth,height=0.25\textheight]{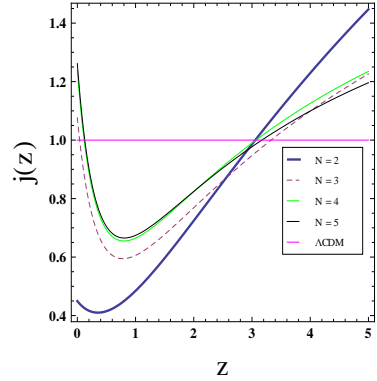}
\caption{\normalsize{\em The evolution of $j(z)$ with respect to redshift $z$ is shown using the best-fit values of $q_{0}$ and $q_{1}$ for different values of $N$, as indicated in the frame. The horizontal thick line (magenta) indicates $j = 1$ (constant) for a $\Lambda$CDM model.}}
\label{figjerk}
\end{figure}

 Figure \ref{figcontourqcsn} shows the $1\sigma$ and $2\sigma$ contours in $q_{0}-q_{1}$ plane for the logarithm parametrization given by equation (\ref{eqdplog2g}). We have found that the best-fit values of the free parameters ($q_{0}$ and $q_{1}$) for the SNIa+BAO/CMB dataset are well fitted in the $1\sigma$ confidence contour.  
%%%%%%%%%%%%%%%%%%%%%%%%%%%%%%%%%%%%%%%%%%%%%%%%%%%%%%%%%%%%%%%%%%%%%%%%%%%%%%%
%%%%%%%%%%%%%%%%%%%%%%%%%%%%%%%%%%%%%%%%%%%%%%%%%%%%%%%%%%%%%%%%%%%%%%%%%%%%%%%%%%% 
%%%%%%%%%%%%%%%%%%%%%%%%%%%%%%%%%%%%%%%%%%%%%%%%%%%%%%%%%%%%%
\begin{table*}
\caption{Results of statistical analysis (within $1\sigma$ confidence level) for Model 1 by considering different values of $N$. Here, $\chi^{2}_{min}$ denotes the minimum value of  $\chi^{2}$. For this analysis we have considered SNIa+BAO/CMB dataset.}
%%%%%%%%%%%%%%%%%%%%%%%%%%%%%%%%%%%%%%%%%%%%%%%%%
{\begin{tabular*}{\textwidth}{@{\extracolsep{\fill}}lrrl@{}}
\hline
$N$&$q_{0}$&$q_{1}$&$\chi^{2}_{min}$\\
\hline
%%%%%%%%%%%%%%%%%%%%%%%%%%%%%%%%%%%%%%%%%%%%%%%%%%%%%%%%%%%%%%%%%%%%%%%%%%%%%
$2$&$-0.45^{+0.07}_{-0.06}$&$-2.56^{+0.27}_{-0.27}$
&$34.69$\\
\hline
%%%%%%%%%%%%%%%%%%%%%%%%%%%%%%%%%%%%%%%%%%%%%%%%%%%%%%%%%%%%%%%%%%%%%%%%%%%%%
$3$&$-0.54^{+0.05}_{-0.04}$&$-1.35^{+0.09}_{-0.09}$
&$34.51$\\
\hline
%%%%%%%%%%%%%%%%%%%%%%%%%%%%%%%%%%%%%%%%%%%%%%%%%%%%%%%%%%%%%%%%%%%%%%%
$4$&$-0.56^{+0.05}_{-0.05}$&$-1.03^{+0.08}_{-0.06}$
&$34.50$\\
\hline
%%%%%%%%%%%%%%%%%%%%%%%%%%%%%%%%%%%%%%%%%%%%%%%%%%%%%%%%%%%%%%%%%%%%%%%%%%%%%
$5$&$-0.56^{+0.04}_{-0.04}$&$-0.85^{+0.05}_{-0.04}$
&$34.17$\\
\hline
%%%%%%%%%%%%%%%%%%%%%%%%%%%%%%%%%%%%%%%%%%%%%%%%%%%%%%%%%%%%%%%%%%%%%%%
\end{tabular*}
\label{table1}}
\end{table*}
%%%%%%%%%%%%%%%%%%%%%%%%%%%%%%%%%%%%%%%%%%%%
%%%%%%%%%%%%%%%%%%%%%%%%%%%%%%%%%%%%%%%%%%%%%%%%%%%%%%%%%%%%%
\begin{table*}
\caption{Results of statistical analysis (within $1\sigma$ confidence level) for Model 2 $\&$ 3 by considering SNIa+BAO/CMB dataset.}
%%%%%%%%%%%%%%%%%%%%%%%%%%%%%%%%%%%%%%%%%%%%%%%%%
{\begin{tabular*}{\textwidth}{@{\extracolsep{\fill}}lrrl@{}}
\hline
Model&$q_{0}$&$q_{1}$&$\chi^{2}_{min}$\\
\hline
%%%%%%%%%%%%%%%%%%%%%%%%%%%%%%%%%%%%%%%%%%%%%%%%%%%%%%%%%%%%%%%%%%%%%%%%%%%%%
Model 2&$-0.41^{+0.03}_{-0.03}$&$0.17^{+0.07}_{-0.07}$
&$32.05$\\
\hline
%%%%%%%%%%%%%%%%%%%%%%%%%%%%%%%%%%%%%%%%%%%%%%%%%%%%%%%%%%%%%%%%%%%%%%%%%%%%%
Model 3&$-0.64^{+0.06}_{-0.05}$&$1.36^{+0.09}_{-0.08}$
&$34.15$\\
\hline
%%%%%%%%%%%%%%%%%%%%%%%%%%%%%%%%%%%%%%%%%%%%%%%%%%%%%%%%%%%%%%%%%%%%%%%
\end{tabular*}
\label{table2}}
\end{table*}
%%%%%%%%%%%%%%%%%%%%%%%%%%%%%%%%%%%%%%%%%%%%
%%%%%%%%%%%%%%%%%%%%%%%%%%%%%%%%%%%%%%%%%%%%%%%%%%%%%%%%%%%%%
\begin{table*}
\caption{Best fit values of $z_{t}$ and $j_{0}$ (within $1\sigma$ errors) for Model 1 by considering different values of $N$. Here, $z_{t}$ is the transition redshift and $j_{0}$ is the present value of $j$. For this analysis we have considered SNIa+BAO/CMB dataset.}
%%%%%%%%%%%%%%%%%%%%%%%%%%%%%%%%%%%%%%%%%%%%%%%%%
{\begin{tabular*}{\textwidth}{@{\extracolsep{\fill}}lrl@{}}
\hline
$N$&$z_{t}$&$j_{0}$\\
\hline
%%%%%%%%%%%%%%%%%%%%%%%%%%%%%%%%%%%%%%%%%%%%%%%%%%%%%%%%%%%%%%%%%%%%%%%%%%%%%
$2$&$1.31^{+0.05}_{-0.06}$&$0.45^{+0.11}_{-0.10}$\\
\hline
%%%%%%%%%%%%%%%%%%%%%%%%%%%%%%%%%%%%%%%%%%%%%%%%%%%%%%%%%%%%%%%%%%%%%%%%%%%%%
$3$&$0.97^{+0.04}_{-0.04}$&$1.07^{+0.12}_{-0.12}$\\
\hline
%%%%%%%%%%%%%%%%%%%%%%%%%%%%%%%%%%%%%%%%%%%%%%%%%%%%%%%%%%%%%%%%%%%%%%%
$4$&$0.89^{+0.04}_{-0.04}$&$1.23^{+0.14}_{-0.15}$\\
\hline
%%%%%%%%%%%%%%%%%%%%%%%%%%%%%%%%%%%%%%%%%%%%%%%%%%%%%%%%%%%%%%%%%%%%%%%%%%%%%
$5$&$0.85^{+0.04}_{-0.03}$&$1.26^{+0.11}_{-0.12}$\\
\hline
%%%%%%%%%%%%%%%%%%%%%%%%%%%%%%%%%%%%%%%%%%%%%%%%%%%%%%%%%%%%%%%%%%%%%%%
\end{tabular*}
\label{table3}}
\end{table*}
%%%%%%%%%%%%%%%%%%%%%%%%%%%%%%%%%%%%%%%%%%%%
%%%%%%%%%%%%%%%%%%%%%%%%%%%%%%%%%%%%%%%%%%%%%%%%%%%%%%%%%%%%%%%%%%%%%%%%%%%%%%%%%%% 
In particular, the $\chi^{2}$ analysis is done by fixing the parameter $N$ to some constant value and the best fit values of $q_{0}$ and $q_{1}$ are presented in table \ref{table1}. Using those best-fit values of $q_{0}$ and $q_{1}$, we have then reconstructed the deceleration parameter $q(z)$ and the results are plotted in figure \ref{figcontourqz}. It is evident from figure \ref{figcontourqz} that $q(z)$ favors the past deceleration $(q>0)$ and recent acceleration $(q<0)$ of the universe. This is essential for the structure formation of the universe.
%%%%%%%%%%%%%%%%%%%%%%%%%%%%%%%%%%%%%%%%%%%%%%%%%%%%%%%%%%%%%%%%%%%%%%%%%%%%%%%
%%%%%%%%%%%%%%%%%%%%%%%%%%%%%%%%%%%%%%%%%%%%%%%%%%%%%%%%%%%%%%%%%%%%%%%%%%%%%%%%%%% 
In this work, we have also obtained the best fit values of the transition redshift ($z_{t}$) within $1\sigma$ errors for the SNIa+BAO/CMB dataset and presented them in table \ref{table3}. This results are found to be consistent with the results obtained by many authors from different considerations \cite{dp10qlog,dp12qlog}. For comparison, the corresponding curves for the Model 2, Model 3 and the $\Lambda$CDM model are also plotted in figure \ref{figcontourqz} which shows that the evolution of $q(z)$ is not always compatible with Model 2 for different choices of $N$. It is also clear from figure \ref{figcontourqz} that the best fit values of $q_{0}$ and $z_{t}$ are in good agreement with the standard $\Lambda$CDM model (within $2\sigma$ errors) and the Model 3 (within $1\sigma$ confidence level) if we increase the value of $N$. From figure \ref{figqfall}, it is seen that as $z$ approaches $-1$, the evolution of $q(z)$ deviates from that of $\Lambda$CDM model. Such a behavior of $q(z)$ may be an outcome of the choice ``$\kappa =\ln N$" made in equation (\ref{eqdplog2g}). As mentioned earlier, the choice of $\kappa =\ln N$ was made such that $q_{0}$ provides the present value of deceleration parameter. Some other choice of $\kappa$ may as well be made so as to make $\Lambda$CDM model consistent in far future and thus needs more detailed analysis.
%%%%%%%%%%%%%%%%%%%%%%%%%%%%%%%%%%%%%%%%%%%%%%
%%%%%%%%%%%%%%%%%%%%%%%%%%%%%%%%%%%%%%%%%%%%%%%%%%%%%%%%%%%%%%%%%%
%%%%%%%%%%%%%%%%%%%%%%%%%%%%%%%%%%%%%%%%%%%%%%
\begin{figure}
\centering
\includegraphics[width=0.32\textwidth,height=0.25\textheight]{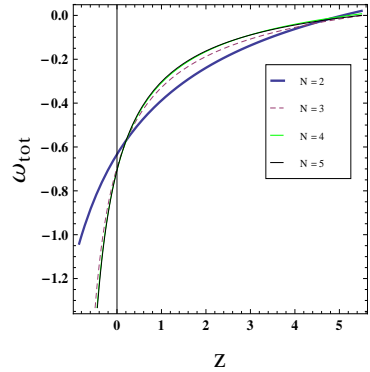}
\caption{\normalsize{\em The reconstructed total EoS parameter $\omega_{tot}(z)$ for the best-fit model by considering different values of $N$, as indicated in the panel (see also Table \ref{table1}).}}
\label{figcontourwsntot}
\end{figure}
%%%%%%%%%%%%%%%%%%%%%%%%%%%%%%%%%%%%%%%%%%%%%%%%%%%%%%%%%%%%%%%%%%
%%%%%%%%%%%%%%%%%%%%%%%%%%%%%%%%%%%%%%%%%%%%%%%%%%%%%%%%%%%%%%%%%%
The best fit evolution of the total EoS parameter $\omega_{tot}$ as a function of $z$, given by equation (\ref{eqomtotal}), is shown in figure \ref{figcontourwsntot}, which indicates that $\omega_{tot}$ attains the required value of $-\frac{1}{3}$ at $z<1$ (or, at $z<1.4$ for $N=2$) and remains greater than $-1$ up to the current epoch (i.e., $z=0$) for the combined (SNIa+BAO/CMB) dataset. This result is consistent with the recent observational results. On the other hand, figure \ref{figjerk} clearly shows the departure of $j$, given by equation (\ref{eqjerkz30}), from the flat $\Lambda$CDM model ($j = 1$) for the best-fit model. It is seen from figure \ref{figjerk} that for the best-fit model, the present value of $j$ is greater than $1$ for $N=3,4$ and $5$, while for $N=2$, $j$ is less than $1$ (see also Table \ref{table3}). So, the present model (with $j\neq 1$ and $q_{0}<0$) clearly indicates that a dynamical DE is more likely to be responsible for the current acceleration.\\
%%%%%%%%%%%%%%%%%%%%%%%%%%%%%%%%%%%%%%%%%%%%%%%%%%%%%%%%%%%%%%%%%%%%%%%%%%%%%%%
%%%%%%%%%%%%%%%%%%%%%%%%%%%%%%%%%%%%%%%%%%%%%%%%%%%%%%%%%%%%%%%%%
\par For statistical comparison of Model 1 with the Models 2 $\&$ 3, two model selection criterion have been used, the Akaike information criterion (AIC) and the Bayesian Information Criterion (BIC). The AIC is defined as \cite{aic}
\be
AIC=-2{\rm ln}L_{max} + 2g 
\ee
and the BIC is defined as \cite{bic}
\be
BIC=-2{\rm ln}L_{max} + g{\rm ln}k 
\ee
where, $L_{max}$ is the maximum likelihood (equivalently, minimum of $\chi^2$) obtained for the model, $g$ is the number of free parameters in that model and $k$ is the number of data points used for the data analysis. If the magnitude of the differences between the AIC ($\triangle$AIC) of the two models (or $\triangle$BIC) is less than 2, then the toy model under consideration (here, Model 1) is strongly favored by the reference model (here, Models 2 $\&$ 3). On the other hand, if the magnitude of $\triangle$AIC or $\triangle$BIC is greater than 10, then the models strongly disfavor each other. As mentioned earlier, the statistical analysis is done by fixing the parameter $N$ to some constant. Because of this, all the three models (Models 1, 2 $\&$ 3) have same number of degrees of freedom, and thus $\triangle$AIC or $\triangle$BIC will be same. Now for Model 1 in comparison with the Model 2 and 3, the $\triangle$AIC (or $\triangle$BIC) values are given by
\be
\triangle AIC=\chi^{2}_{min}({\rm Model}~1) - \chi^{2}_{min}({\rm Model}~2)\\ \nonumber
\ee
and
\be
\triangle AIC=\chi^{2}_{min}({\rm Model}~1) - \chi^{2}_{min}({\rm Model}~3)\\ \nonumber
\ee
Now, it is clear from table \ref{table1} and \ref{table2} that for the model 3, the $\triangle$AIC (or $\triangle$BIC) is less than 2 for each choices of $N$. Hence, the present reconstructed model is highly consistent with the model 3.
%%%%%%%%%%%%%%%%%%%%%%%%%%%%%%%%%%%%%%%%%%%%%%%%%%%%%%%%%%%%%%%%%%%%%
\section{Conclusions}\label{conclusa}
%%%%%%%%%%%%%%%%%%%%%%%%%%%%%%%%%%%%%%%%%%%%%%%%%%%%%%%%%%%%
In the present work, we have studied the dynamics of accelerating scenario within the framework of scalar field model. Here, we have considered one specific parameterization of the deceleration parameter $q(z)$ and from this we have found out analytical solutions for various cosmological parameters. As we have seen before, the new parametrization of $q(z)$ is similar to the well-known parametrization of $q(z)$ for appropriate choices of $q_{0}$, $q_{1}$ and $N$. We have also compared our theoretical model with three different popular models, such as $q\propto z$ (see equation (\ref{eqmodel2})), $q\propto \frac{z}{1+z}$(see equation (\ref{eqmodel3})) and $\Lambda$CDM (see equation (\ref{eqmodellcdm})), to draw a direct comparison between them. In what follows, we have summarized the main results of our analysis. 
%%%%%%%%%%%%%%%%%%%%%%%%%%%%%%%%%%%%%%%%%%%%%%%%%%%%%%%%%%%%%%%%%%%%%% 
\par The observational data analysis by $\chi^{2}$-minimization technique have also been analyzed for this model using the SNIa+BAO/CMB dataset. From this analysis, we have obtained the bounds on the arbitrary parameters $q_{0}$ and $q_{1}$ within $1\sigma$ and $2\sigma$ confidence levels. It has been found that $q(z)$ shows exactly the behavior which is desired, a deceleration for high $z$ limit whereas an acceleration for the low $z$ limit. This is essential to explain both these observed growth of structure at the early epoch and the late-time cosmic acceleration measurements. It should be noted that for the present model, the values of the transition redshift $z_{t}$ with $1\sigma$ errors are consistent with the values obtained by many authors from different scenarios \cite{dp10qlog,dp12qlog}.  For this model, the jerk parameter $j$ is found to be evolving, which indicates a tendency of deviation of the universe from the standard $\Lambda$CDM model. These results  make the present work worthy of attention. The present model also provides an analytical solution for the EoS parameter $\omega_{\phi}(z)$. Interestingly, it has been found that the EoS parameter reduces to the well-known CPL parametrization of $\omega_{\phi}(z)$ for low $z$ \cite{cpl1,cpl2}. To obtain more physical insight regarding the evolution of $\omega_{\phi}(z)$, we have plotted the reconstructed total EoS parameter in figure \ref{figcontourwsntot} and the resulting  scenarios agree very well with the observational results at the present epoch. 
%%%%%%%%%%%%%%%%%%%%%%%%%%%%%%%%%%%%%%%%%%%%%%%%%%
\par As discussed earlier, this particular choice of $q(z)$ is quite arbitrary and we have made this assumption to close the system of equations. Since the nature of the universe is still a mystery, therefore the idea to parameterize $q(z)$ is a simple approach to study the transition of the universe from decelerated to accelerated expansion phase and also opens up possibilities for future studies regarding the nature of dark energy.  Definitely, the addition of more observational datasets in the present work may help us to obtain more precise constraints on the expansion history of the universe and the present work is one preliminary step towards that direction.
%%%%%%%%%%%%%%%%%%%%%%%%%%%%%%%%%%%%%%%%%%%%%%%%%%%%%
\section{Acknowledgements}
The authors are thankful to the anonymous referee whose useful suggestions have improved the quality of the paper. A.A.M. acknowledges UGC, Govt. of India, for financial support through a Maulana Azad National Fellowship. SD wishes to thank IUCAA, Pune for associateship program.
%%%%%%%%%%%%%%%%%%%%%%%%%%%%%%%%%%%%%%%%%%%%%%%%%%%%%%%%%%%%%%%%%%%%%%%%%% 

\end{document}